\documentstyle[pra,aps,preprint]{revtex}
\begin{document}
\draft
\title{QED-SCF, MCSCF and Coupled-cluster Methods in Quantum Chemistry}
\author{Tadafumi Ohsaku\footnote{Corresponding author; Tadafumi Ohsaku, tadafumi@fuji.phys.wani.osaka-u.ac.jp} and Kizashi Yamaguchi}
\address{Department of Physics and Department of Chemistry, Graduate School of Science, Osaka University, Machikaneyama-cho 1-1, Toyonaka, Osaka, 560-0043, Japan}

\maketitle

\begin{abstract}

We investigate the method to combine the techniques of quantum chemisty with QED. In our theory, we treat the N-electron system and the Dirac sea on an equal footing; we regard both of them as the dynamical degrees of freedom of a many-body system. After the introduction of our QED-SCF method, the QED-SCF solutions are classified into several classes on the basis of group-theoretical operations such as time reversal, parity and $O(3)$ rotational symmetry. The natural orbitals of general QED-SCF solutions are determined by diagonalizing the first order density matrix. Thus, we obtain the possibility to treat the system under strong QED effect by the methods of quantum chemistry, such as QED-MCSCF and QED-coupled-cluster approaches.

\end{abstract}

keywords; QED, QED-SCF, MCSCF, Coupled-cluster theory, Radiation chemistry. 

\section{Introduction}

The Hartree-Fock ( HF ) and multiconfiguration HF ( MCHF ) methods have been well-established for nonrelativistic Hamiltonian, and the self-consistent field ( SCF ) equations to obtain the HF and MCHF solutions are also derived in several ways~[1]. The most general HF solution for the nonrelativistic case is given by the general spin orbitals ( GSO ): The two-component spinor~[2,3],
\begin{equation}
\phi_{i} = \phi^{+}_{i}\alpha + \phi^{-}_{i}\beta = \left(
\begin{array}{c}
\psi^{+}_{i} \\
\psi^{-}_{i}
\end{array}
\right).
\end{equation}
Very recently we have developed the program package for ab initio HF and density functional ( DFT ) calculations of molecules by using the GSO~[4,5]. It was shown that GSO HF and GSO DFT approaches are useful for qualitative understanding of correlation and spin correlation effects in polyradicals with orbital degeneracies~[4,5].

By the way, it is a well-known fact that, in atomic and moleculer physics, the relativistic effects become crucial in many cases. Especially in the electronic structure calculation for heavy elements, we have to introduce a kind of relativistic treatment. For example, the case of neutral heavy elements  can be treated satisfactorily by the Dirac-Coulomb-Breit no-sea scheme~[6]. This method is based on the no-sea (the approximation completely neglects the effects of the Dirac sea) Hamiltonian. It has a Dirac-type one-body Hamiltonian, and the interaction of electrons is treated via the Coulomb potential and Breit operator~[7]. This Hamiltonian with several many-body schemes (HF, relativistic many-body perturbation theory ( RMBPT ), so on) give good numerical results for neutral heavy elements in atoms and molecules. But, especially in the case of highly charged heavy elements, the accurate second-order RMBPT indicates that quantum electrodynamical ( QED ) corrections are clearly seen in the difference between RMBPT and experiment. Various relativistic schemes have already been presented in the electronic structure theory,
and they are summarized in recent review articles~[6,8,9].

In the previous paper~[10], we constructed the theory
of quantum electrodynamical self-consistent fields ( QED-SCF ).
We derived the time-dependent Hartree-Fock ( TDHF ) theory, HF condition and random phase approximation ( RPA )~[11,12]. Because we constructed our theory based on QED, we regard them as an intrinsic treatment of relativistic theory in atomic and moleculer physics. In this theory, we treat an N-electron system and the Dirac sea on an equal footing, and they interact by exchanging photons. Thus our theory can treat the QED effects such as the vacuum polarization effect, which will be observed around the nucleus of heavy element ( strong external Coulomb field )~[8,9,13]. To describe the QED effects, we have to treat not only N-electron system but also the Dirac sea as the dynamical degrees of freedom in a many-body system. Futhermore, because we gain the ability to introduce various many-body techniques of atoms and molecules to QED, we can discuss the availabilities of post-HF scheme through consideration of the QED-HF stability condition, like the GSO model in the nonrelativistic theory.  

This paper is organized as follows. In Sec. II, we give a brief summary of our method to combine many-body techniques to QED. In Sec. III, we perform group-theoretical characterizations of QED-SCF solutions~[10] on the basis of time-reversal, parity and $O(3)$ rotational symmetry. Under these group-operations, the first-order density matrix of these solutions is classified into several different classes. In Sec. IV, we give some examples to apply our method to quantum chemistry. We examine the post QED-SCF methods as in the case of post HF methods in quantum chemistry. The MBPT, coupled-cluster ( CC ) and MCSCF methods are considered in the QED scheme. To this end, we utilize previous procedures for the nonrelativistic case~[14]: (1) Use of density matrix, (2) its diagonalization to obtain the natural orbitals given by four-component bispinor, and (3) selection of active space for MCSCF. Finally in Sec. V, we give discussion about this work.

\section{QED-HF Theory}

In this section, we give a brief summary of our QED-SCF method which has been 
investigated in previous paper~[10]. Our QED-SCF method is based on three
factors: First, we introduce a QED Hamiltonian written by products of the
one-particle operators ( creation-annihilation operators ).
Second, we introduce an assumption for evaluation of matrix elements.
Third, we use the Thouless form for the determinantal state~[11,15].
By introducing these three tools, it becomes possible to combine various
many-body techniques with QED. 

\subsection{Hamiltonian}

We start from the following Lagrangian
\begin{eqnarray}
{\cal L}(x) &=& -\frac{1}{4}F_{\mu\nu}(x)F^{\mu\nu}(x) + \bar{\psi}(x)(i\gamma^{\mu}\partial_{\mu}-m_{0})\psi(x) +e\bar{\psi}(x)\gamma^{\mu}\psi(x)(A_{\mu}(x)+A^{(e)}_{\mu}(x)).
\end{eqnarray}
Here, $A_{\mu}$ is a ( virtual ) photon field which describes the interaction
between fermions, and $A^{(e)}_{\mu}$ is a classical external field.
Next we remove the photon degree of freedom and introduce 
a Coulomb potential $U(r)$ of a nucleus ( or nuclei ) as the external field.
Then we obtain the following Hamiltonian:
\begin{eqnarray}
\hat{H} &=& \int d^{3}x\hat{\bar{\psi}}(x)(-i\vec{\gamma}\cdot\nabla+m_{0}+\gamma^{0}U(r))\hat{\psi}(x) \nonumber \\
  & & +\frac{1}{2}e^{2}\int d^{3}x\int d^{4}y \hat{\bar{\psi}}(x)\gamma^{\mu}\hat{\psi}(x)D_{\mu\nu}(x-y)\hat{\bar{\psi}}(y)\gamma^{\nu}\hat{\psi}(y). 
\end{eqnarray}
Here each field is quantized. $D_{\mu\nu}(x-y)$ is the full photon propagator.
The $\gamma$-matrices are defined by ( under the standard representation )
\begin{equation}
\gamma^{0}=\left(
\begin{array}{cc}
I & 0 \\
0 & -I 
\end{array}
\right), \qquad \gamma^{i}=\left(
\begin{array}{cc}
0 & \sigma^{i} \\
-\sigma^{i} & 0
\end{array}
\right),
\end{equation}
where $I$ is the unit matrix, and $\sigma_{i}$ ( i=1,2,3 ) is the Pauli matrix:
\begin{equation}
\sigma^{1}=\left(
\begin{array}{cc}
0 & 1 \\
1 & 0 
\end{array}
\right), \qquad \sigma^{2}=\left(
\begin{array}{cc}
0 & -i \\
i & 0 
\end{array}
\right), \qquad \sigma^{3}=\left(
\begin{array}{cc}
1 & 0 \\
0 & -1 
\end{array}
\right).
\end{equation}
In the above Hamiltonian, we expand each field operator by one-particle state
function as
\begin{eqnarray}
\hat{\psi}(x) &=& \sum_{i}(\psi^{(+)}_{i}(x)\hat{a}_{i}+\psi^{(-)}_{i}(x)\hat{b}^{\dagger}_{i}), \\
\hat{\bar{\psi}}(x) &=& \sum_{i}(\bar{\psi}^{(+)}_{i}(x)\hat{a}^{\dagger}_{i}+\bar{\psi}^{(-)}_{i}(x)\hat{b}_{i}),  
\end{eqnarray}
where $+(-)$ means the electron ( positron ) state, $i$ denotes quantum numbers for one-particle states, $\hat{a}^{\dagger}_{i}(\hat{a}_{i})$ is electron creation ( annihilation ) operator, $\hat{b}^{\dagger}_{i}(\hat{b}_{i})$ is positron creation ( annihilation ) operator, and $\psi^{(\pm)}_{i}$ is the four-component bispinor given like
\begin{eqnarray}
\psi^{(\pm)}_{i}=(\chi^{(\pm)(1)}_{i},\chi^{(\pm)(2)}_{i},\chi^{(\pm)(3)}_{i},\chi^{(\pm)(4)}_{i}).
\end{eqnarray} 
Then we obtain the Hamiltonian written by creation-annihilation:
operators
\begin{eqnarray}
\hat{H} &=& \hat{H}_{K} + \hat{H}_{I},  \\
\hat{H}_{K} &=& \sum_{i,j} T^{\pm\pm}_{ij}( \hat{a}^{\dagger}_{i}\hat{a}_{j} + \hat{a}^{\dagger}_{i}\hat{b}^{\dagger}_{j} + \hat{b}_{i}\hat{a}_{j} - \hat{b}^{\dagger}_{j}\hat{b}_{i} ), \\
\hat{H}_{I} &=& \frac{1}{2} \sum_{i,j,k,l} V^{\pm\pm\pm\pm}_{ijkl}(\hat{a}^{\dagger}_{i}\hat{a}^{\dagger}_{k}\hat{a}_{l}\hat{a}_{j} + \hat{a}^{\dagger}_{i}\hat{a}^{\dagger}_{k}\hat{b}^{\dagger}_{l}\hat{a}_{j} + \hat{a}^{\dagger}_{i}\hat{a}_{j}\hat{b}_{k}\hat{a}_{l} + \hat{a}^{\dagger}_{i}\hat{b}^{\dagger}_{l}\hat{a}_{j}\hat{b}_{k} \nonumber \\
 & & + \hat{a}^{\dagger}_{i}\hat{b}^{\dagger}_{j}\hat{a}^{\dagger}_{k}\hat{a}_{l} + \hat{a}^{\dagger}_{i}\hat{b}^{\dagger}_{j}\hat{a}^{\dagger}_{k}\hat{b}^{\dagger}_{l} + \hat{a}^{\dagger}_{i}\hat{b}^{\dagger}_{j}\hat{b}_{k}\hat{a}_{l} + \hat{a}^{\dagger}_{i}\hat{b}^{\dagger}_{l}\hat{b}^{\dagger}_{j}\hat{b}_{k} + \hat{a}^{\dagger}_{k}\hat{b}_{i}\hat{a}_{j}\hat{a}_{l} + \hat{a}^{\dagger}_{k}\hat{b}^{\dagger}_{l}\hat{b}_{i}\hat{a}_{j} \nonumber \\
 & & + \hat{b}_{i}\hat{a}_{j}\hat{b}_{k}\hat{a}_{l} + \hat{b}^{\dagger}_{l}\hat{a}_{j}\hat{b}_{i}\hat{b}_{k} + \hat{b}^{\dagger}_{j}\hat{a}^{\dagger}_{k}\hat{b}_{i}\hat{a}_{l} + \hat{a}^{\dagger}_{k}\hat{b}^{\dagger}_{j}\hat{b}^{\dagger}_{l}\hat{b}_{i} + \hat{b}^{\dagger}_{j}\hat{b}_{k}\hat{b}_{i}\hat{a}_{l} + \hat{b}^{\dagger}_{j}\hat{b}^{\dagger}_{l}\hat{b}_{k}\hat{b}_{i}), 
\end{eqnarray}
where we designate matrix elements as
\begin{eqnarray}
T^{\pm\pm}_{ij} &=& \int d^{3}x \bar{\psi}^{(\pm)}_{i}({\bf x})(-i\vec{\gamma}\cdot\nabla + m_{0} + \gamma_{0}U(r))\psi^{(\pm)}_{j}({\bf x}),
\end{eqnarray}
\begin{eqnarray}
V^{\pm\pm\pm\pm}_{ijkl} &=& \frac{1}{2}e^{2}\int d^{3}x \int d^{3}y \bar{\psi}^{(\pm)}_{i}({\bf x})\gamma^{\mu}\psi^{(\pm)}_{j}({\bf x})\frac{g_{\mu\nu}\exp(i\Delta\epsilon|{\bf x}-{\bf y}|)}{4\pi|{\bf x}-{\bf y}|} \bar{\psi}^{(\pm)}_{k}({\bf y})\gamma^{\nu}\psi^{(\pm)}_{l}({\bf y}).  \nonumber \\
& & 
\end{eqnarray}
Here, $\Delta\epsilon = |\epsilon_{k}-\epsilon_{l}| = |\epsilon_{i}-\epsilon_{j}|$ ( the energy difference of  one-particle states ). $g_{\mu\nu}$ is usual metric tensor and defined as $g_{\mu\nu}={\rm diag}(1,-1,-1,-1)$. About the interaction, we only take into account 0th order bare photon propagator, and choose the Feynman gauge for convenience. This Hamiltonian can describes the QED effects such as the vacuum-polarization, because which includes the Dirac sea as the dynamical degree of freedom. Hereafter we use this Hamiltonian in our theory. 

Next we give the method for evaluation of expectation value. 
In QED, the Dirac vacuum under the presence of the external Coulomb field generates 4-current as an observed effect, which is called the vacuum polarization~[8,9,13]. It is well-known that, this effect is described by the charge conjugation symmetric 4-current given in the next form~[8,9,13]
\begin{eqnarray}
j^{vac}_{\mu}(x) &=& -\frac{e}{2} \biggl( \sum_{n<-m} \bar{\psi}^{(-)}_{n}(x)\gamma_{\mu}\psi^{(-)}_{n}(x) - \sum_{n>-m} \bar{\psi}^{(+)}_{n}(x)\gamma_{\mu}\psi^{(+)}_{n}(x) \biggr).
\end{eqnarray}
This expression gives the 4-current induced by the external Coulomb field. In this expression, in the external-field-free case two sums cancel, but for the field present case two sums do not cancel completely. Hereby, this result should be interpreted as describing the vacuum polarization effect which is induced by the external Coulomb field. Therefore, the current of N-electron system with adding the vacuum polarization ( here we consider the case that the N-electrons occupy the one-particle states up to the Fermi level, as illustrated in Fig. 1 ) is given as
\begin{eqnarray}
j_{\mu}(x) &=& -e\sum_{-m<i\le F}\bar{\psi}^{(+)}_{i}(x)\gamma_{\mu}\psi^{(+)}_{i}(x) \nonumber \\
 & & -\frac{e}{2}\Biggl( \sum_{i<-m}\bar{\psi}^{(-)}_{i}(x)\gamma_{\mu}\psi^{(-)}_{i}(x) - \sum_{i>-m}\bar{\psi}^{(+)}_{i}(x)\gamma_{\mu}\psi^{(+)}_{i}(x)\Biggl).  
\end{eqnarray}
The second line gives the vacuum polarization~[8,9,13]. To include this effect in our Hamiltonian, we have to modify the contraction scheme. From the definition of the contraction: 
\begin{eqnarray}
A^{\bullet}(t)B^{\bullet}(t') &=& \langle F|A(t)B(t')|F \rangle \theta(t-t') - \langle F|B(t')A(t)|F \rangle \theta(t'-t). 
\end{eqnarray}
Here, $A(t)$ and $B(t)$ are some kind of operators, and $|F\rangle$ is a Fermi sea. We introduce the definition for the step function at the same time as $\theta(t-t')_{t=t'} = \theta(t'-t)_{t=t'} = \frac{1}{2}$. Then we obtain the following relation:
\begin{eqnarray}
\langle F|T(A(t)B(t'))|F \rangle _{t=t'} &=& \langle F|A^{\bullet}(t)B^{\bullet}(t')|F \rangle _{t=t'} \nonumber \\
 &=& \frac{1}{2}\langle F|A(t)B(t)|F \rangle -\frac{1}{2}\langle F|B(t)A(t)|F \rangle. 
\end{eqnarray}
Let us consider the fact that, it is clear from the Hamiltonian given in Eq. (3), the expectation value of it will be written in {\it a functional of the 4-current} $j_{\mu}$, as same as the discussion in Ref. 8. Therefore, we introduce a hypothesis that our Hamiltonian is written by operators in the same-time $T$-products, and when we factorize the vacuum expectation value for an operator product with the aid of the Wick theorem, each contraction should be calculated by the definition given in (17).
 
By using the above method, for example, the expectation value of $\hat{H}$ is
given by 
\begin{eqnarray}
\langle F|\hat{H}|F \rangle &=& \frac{1}{2}\sum_{i<-m} T^{--}_{ii} + \frac{1}{2}(\sum_{-m<i \le F} - \sum_{i>F} )T^{++}_{ii} \nonumber \\
&+&  \frac{1}{8}\biggl\{ (\sum_{-m<i\le F}-\sum_{i>F})(\sum_{-m<k \le F}-\sum_{k>F})(V^{++++}_{iikk}-V^{++++}_{ikki}) \nonumber \\
 & & +(\sum_{-m<i \le F}-\sum_{i>F})(\sum_{k<-m})(V^{++--}_{iikk}-V^{+--+}_{ikki}) \nonumber \\
 & & +(\sum_{-m<k\le F}-\sum_{k>F})(\sum_{i<-m})(V^{--++}_{iikk}-V^{-++-}_{ikki}) \nonumber \\
 & & +(\sum_{i<-m})(\sum_{k<-m})(V^{----}_{iikk}-V^{----}_{ikki}) \biggr\}.
\end{eqnarray}
This gives the HF energy in our theory. We argue that we can generalize this method to calculate expectation values for arbitrary products of operators. By using the Wick theorem, we always factorize the matrix elements of an operator product into the sum of the products of contractions, then the contraction is calculated by the definition given above. It is clear from the above discussion that {\it only} the definition of the contraction for same-time operators is modified.

\subsection{HF solution}

Now, to obtain the relativistic Slater determinant in the Thouless form~[11], we will obtain a relativistic exponential operator which transforms one to another representations.
We assume the following relations for the canonical transformation:
\begin{eqnarray}
\hat{a}'_{m} &=& e^{i\hat{S}_{1}}\hat{a}_{m}e^{-i\hat{S}_{1}}  =  \sum_{n}(\alpha_{mn}\hat{a}_{n} + \beta_{mn}\hat{b}^{\dagger}_{n}), \\
\hat{a}'^{\dagger}_{m} &=& e^{i\hat{S}_{1}}\hat{a}^{\dagger}_{m}e^{-i\hat{S}_{1}} = \sum_{n}(\alpha^{*}_{mn}\hat{a}^{\dagger}_{n} + \beta^{*}_{mn}\hat{b}_{n}), \\
\hat{b}'^{\dagger}_{m} &=& e^{i\hat{S}_{1}}\hat{b}^{\dagger}_{m}e^{-i\hat{S}_{1}} = \sum_{n}(\beta_{mn}\hat{a}_{n} + \gamma_{mn}\hat{b}^{\dagger}_{n}), \\
\hat{b}'_{m} &=& e^{i\hat{S}_{1}}\hat{b}_{m}e^{-i\hat{S}_{1}} = \sum_{n}(\beta^{*}_{mn}\hat{a}^{\dagger}_{n} + \gamma^{*}_{mn}\hat{b}_{n}), 
\end{eqnarray}
where $\hat{S}^{\dagger}_{1}=\hat{S}_{1}$ will be satisfied. Here new operators are expanded by a complete set of the old representation. We introduce the exponential operator in the following form:
\begin{eqnarray}
e^{i\hat{S}_{1}} &=& \exp\Bigl\{i\sum_{mn}(\alpha_{mn}\hat{a}^{\dagger}_{m}\hat{a}_{n} + \beta_{mn}\hat{a}^{\dagger}_{m}\hat{b}^{\dagger}_{n} + \beta^{*}_{mn}\hat{b}_{n}\hat{a}_{m} + \gamma^{*}_{mn}\hat{b}_{n}\hat{b}^{\dagger}_{m})\Bigr\}.
\end{eqnarray}
Here we demand that the matrices formed by the paramerters $\alpha_{mn},\beta_{mn},\beta^{*}_{mn},\gamma^{*}_{mn}$ should be Hermitian. 
Then the exponential operator formally given above is unitary. 
It is clear that the operator (23) will give the relations in (19)$\sim$(22). We will write a relativistic Slater determinant as
\begin{eqnarray}
|\Phi(\alpha,\beta,\beta^{*},\gamma^{*})\rangle = e^{i\hat{S}_{1}(\alpha,\beta,\beta^{*},\gamma^{*})}|\Phi_{0}\rangle .
\end{eqnarray}
Here the Slater determinant of old representation is expressed by
\begin{eqnarray}
|\Phi_{0}\rangle &=& \prod_{i}\hat{a}^{\dagger}_{i}|0\rangle 
\end{eqnarray}
and the Dirac vacuum is defined by $\hat{a}_{i}|0\rangle =0$ and $\hat{b}_{i}|0\rangle =0$. 

By using the relativistic Slater determinant given above, 
we can write the expectation value of our Hamiltonian
into an expanded form, by the same way as nonrelativistic cases. 
The expectation value of our Hamiltonian is given as follows:
\begin{eqnarray}
\langle \Phi(\alpha,\beta,\beta^{*},\gamma^{*})|\hat{H}|\Phi(\alpha,\beta,\beta^{*},\gamma^{*})\rangle &=& \langle \Phi_{0}|\hat{H}|\Phi_{0} \rangle \nonumber \\ 
 & & + i\langle \Phi_{0}|[\hat{H},\hat{S}_{1}(\alpha,\beta,\beta^{*},\gamma^{*})]|\Phi_{0}\rangle  \nonumber \\
 & & + \frac{i^{2}}{2!}\langle\Phi_{0}|[[\hat{H},\hat{S}_{1}(\alpha,\beta,\beta^{*},\gamma^{*})],\hat{S}_{1}(\alpha,\beta,\beta^{*},\gamma^{*})]|\Phi_{0}\rangle \nonumber \\
 & &  + {\cal O}(\hat{S}^{3}_{1}). 
\end{eqnarray}
In (26), the first line is the HF total energy, the second line gives the first derivatives with respect to the parameters ($\alpha,\beta,\beta^{*},\gamma^{*}$), and it must be zero in the HF condition. The third line corresponds
to the second derivatives and they determine the stability of the HF state:
\begin{eqnarray}
\frac{i^{2}}{2!}\langle\Phi_{0}|[[\hat{H},\hat{S}_{1}(\alpha,\beta,\beta^{*},\gamma^{*})],\hat{S}_{1}(\alpha,\beta,\beta^{*},\gamma^{*})]|\Phi_{0}\rangle \ge 0,
\end{eqnarray}
and it is equivalent to the stability of collective modes in the RPA. 
It is obvious fact that {\it the operator formalism makes possible to do these discussions in QED}.
We gave the evidence in previous paper~[10] that, by using our method, 
we can derive the TDHF equation, HF condition and RPA equation in QED with no inconsistency.  
Thus we can obtain the most stable generalized QED solution at the HF level.

\section{Group-theoretical classification of generalized QED-HF solutions}

\subsection{Group-theoretical classification}

In this section, we give a brief discussion of group-theoretical classification of the generalized QED-HF solutions. We firstly introduce the density matrix like the work of Fukutome~[16]:
\begin{equation}
Q_{4\times4} = -\langle\psi(x)\bar{\psi}(x)\rangle_{4\times4}.
\end{equation}  
Here $\psi$ and $\bar{\psi}$ are usual Dirac field, and they are 4-component bispinors. Thus our density matrix is 4$\times$4-matrix, as denoted above.
We can expand the 4$\times$4 density matrix into the 16-dimensional complete set of $\gamma$-matrices:
\begin{eqnarray}
Q_{4\times4} &=& Q^{S}I+Q^{V}_{\mu}\gamma^{\mu}+Q^{T}_{\mu\nu}\sigma^{\mu\nu}+Q^{A}_{\mu}\gamma_{5}\gamma^{\mu}+Q^{P}i\gamma_{5}.
\end{eqnarray}
In this expansion, we take a convention that $S$ denotes the scalar, $V$ denotes the vector, $T$ denotes the 2-rank antisymmetric tensor, $A$ denotes the axial vector and $P$ denotes the pseudoscalar. $I$ is the $4\times4$ unit-matrix, $\gamma^{\mu}$ is usual Dirac gamma matrix, $\sigma^{\mu\nu}$ is defined as $\sigma^{\mu\nu}=\frac{i}{2}[\gamma^{\mu},\gamma^{\nu}]$, and $\gamma_{5}=i\gamma^{0}\gamma^{1}\gamma^{2}\gamma^{3}$. Thus we obtain the Lorentz structure in our density matrix $Q$. 

In the relativistic theory, we usually treat the Poincar\'{e} group ( 4-translation and the Lorentz group ), C ( charge conjugation ), P ( parity ), T ( time-reversal ). Under the charge conjugation, $\psi$ and $\bar{\psi}$ are transformed as
\begin{eqnarray}
\psi \to C\bar{\psi}^{T}, \qquad \bar{\psi} \to -\psi^{T}C^{-1}.
\end{eqnarray}
Then the density matrix is transformed as
\begin{eqnarray}
Q &=& -\langle\psi\bar{\psi}\rangle \to -C\langle\psi\bar{\psi}\rangle ^{T}C^{-1}= CQ^{T}C^{-1} \nonumber \\
&=& Q^{S}I-Q^{V}_{\mu}\gamma^{\mu}-Q^{T}_{\mu\nu}\sigma^{\mu\nu}+Q^{A}_{\mu}\gamma_{5}\gamma^{\mu}+Q^{P}i\gamma_{5}.
\end{eqnarray}
Here $C\equiv i\gamma^{2}\gamma^{0}$ is the charge conjugation matrix, and $T$ denotes the transposition of matrix. Under the time reversal,
\begin{eqnarray}
\psi(t) \to T\psi(-t), \qquad \bar{\psi}(t) \to \bar{\psi}(-t)T,
\end{eqnarray}
together with the rule of taking the complex conjugate about c-numbers, we obtain
\begin{eqnarray}
Q(t) &=& -\langle\psi(t)\bar{\psi}(t)\rangle \to -T\langle\psi(-t)\bar{\psi}(-t)\rangle ^{*}T =TQ(-t)^{*}T \nonumber \\
&=& Q^{S*}(-t)I+Q^{V*}_{0}(-t)\gamma^{0}-Q^{V*}_{i}(-t)\gamma^{i}+Q^{T*}_{0i}(-t)\sigma^{0i} \nonumber \\
& & -Q^{T*}_{ij}(-t)\sigma^{ij}+Q^{A*}_{0}(-t)\gamma_{5}\gamma^{0}-Q^{A*}_{i}(-t)\gamma_{5}\gamma^{i}-Q^{P*}(-t)i\gamma_{5}.
\end{eqnarray}
Here $T\equiv i\gamma^{1}\gamma^{3}$ and $i,j=1,2,3$.
Under the spatial inversion,
\begin{eqnarray}
\psi({\bf x}) \to \gamma^{0}\psi(-{\bf x}), \qquad \bar{\psi}({\bf x}) \to \bar{\psi}(-{\bf x})\gamma^{0},
\end{eqnarray}
the density matrix is transformed as
\begin{eqnarray}
Q({\bf x}) &=& -\langle\psi({\bf x})\bar{\psi}({\bf x})\rangle \to -\gamma^{0}\langle\psi(-{\bf x})\bar{\psi}(-{\bf x})\rangle\gamma^{0} =\gamma^{0}Q(-{\bf x})\gamma^{0}\nonumber \\
&=& Q^{S}(-{\bf x})I+Q^{V}_{0}(-{\bf x})\gamma^{0}-Q^{V}_{i}(-{\bf x})\gamma^{i}-Q^{T}_{0i}(-{\bf x})\sigma^{0i}\nonumber \\
& & +Q^{T}_{ij}(-{\bf x})\sigma^{ij}-Q^{A}_{0}(-{\bf x})\gamma_{5}\gamma^{0}+Q^{A}_{i}(-{\bf x})\gamma_{5}\gamma^{i}-Q^{P}(-{\bf x})i\gamma_{5}. 
\end{eqnarray}

The QED-HF solutions can be group-theoretically classified into several types as in the case of nonrelativistic HF solutions~[16]. To consider this problem, we determine the symmetry-group of a system. In the atomic or molecular systems, the translation invariance is broken. In the case of an atom, only $O(3)$ rotation is remained in the Lorentz group ( In the case of a molecule, $O(3)$ is replaced by the point group.). Under the $O(3)$ rotational symmetry, we expand the $Q^{S}$, $Q^{P}$, $Q^{V}_{0}$ and $Q^{A}_{0}$ by the scalar spherical harmonics, while we expand the $Q^{V}_{i}$, $Q^{A}_{i}$ and $Q^{T}_{\mu\nu}$ by the vector spherical harmonics~[17]. It is clear from (33) and (35), the behavior of each type of the density matrix given in (29) under the spatial inversion and time reversal depends not only on the structure of the $\gamma$-matrix, but also on the angular momentum of the spherical harmonics. Let us consider the case of an atom. We treat the group $G=O(3)\times P\times T$. Here, we introduce the subgroup of $G$ as
\begin{eqnarray}
& & O(3)\times P\times T, \quad O(3)\times P, \quad O(3)\times T, \nonumber \\
& & P\times T, \quad O(3), \quad P, \quad T, \quad 1. 
\end{eqnarray}
For example, $O(3)\times P\times T$-invariant solution is given as
\begin{eqnarray}
Q^{S*}=Q^{S},\quad Q^{V*}_{0}=Q^{V}_{0},\quad {\rm others}=0.
\end{eqnarray}
There is no case for $O(3)\times P$-invariant solution.
$O(3)\times T$-invariant solution is given as
\begin{eqnarray}
Q^{S*}=Q^{S},\quad Q^{V*}_{0}=Q^{V}_{0},\quad Q^{A*}_{0}=Q^{A}_{0},\quad {\rm others}=0.
\end{eqnarray}
$O(3)$-invariant solution is given as
\begin{eqnarray}
Q^{S*}=Q^{S},\quad Q^{V*}_{0}=Q^{V}_{0},\quad Q^{A*}_{0}=Q^{A}_{0},\quad Q^{P*}=Q^{P},\quad {\rm others}=0.
\end{eqnarray}
Due to the rotational symmetry, each density matrix can only have an s-wave component, in all cases (37)$\sim$(39). All solutions given above are for closed shell states. We may have magnetic QED-HF solutions for systems under discussions, if time reversal symmetry is broken ( For example, the cases of (38) and (39) ).

\subsection{Nonrelativistic limit}

To consider the nonrelativistic limit, we take the standard representation. The four component spinor $\psi$ is partitioned as~[17]
\begin{equation}
\psi=\left(
\begin{array}{c}
\phi \\
\chi
\end{array}
\right),
\end{equation}
where $\phi$ is the large component and $\chi$ is the small component. Then,  
\begin{eqnarray}
Q^{S} &=& \frac{1}{4}{\rm tr} Q = \frac{1}{4}\langle\bar{\psi}\psi\rangle =\frac{1}{4}\langle\phi^{\dagger}\phi-\chi^{\dagger}\chi\rangle, \\
Q^{V}_{0} &=& \frac{1}{4}{\rm tr} \gamma^{0}Q = \frac{1}{4}\langle\bar{\psi}\gamma^{0}\psi\rangle =\frac{1}{4}\langle\phi^{\dagger}\phi+\chi^{\dagger}\chi\rangle, \\
Q^{A}_{0} &=& \frac{1}{4}{\rm tr} \gamma_{5}\gamma^{0}Q = \frac{1}{4}\langle\bar{\psi}\gamma_{5}\gamma^{0}\psi\rangle =-\frac{1}{4}\langle\chi^{\dagger}\phi+\phi^{\dagger}\chi\rangle, \\
Q^{P} &=& \frac{1}{4}{\rm tr} i\gamma_{5} Q = \frac{1}{4}\langle\bar{\psi}i\gamma_{5}\psi\rangle =\frac{1}{4}\langle-i\chi^{\dagger}\phi+i\phi^{\dagger}\chi\rangle. 
\end{eqnarray}
Therfore, at the nonrelativistic limit, the $Q^{S}$ and $Q^{V}_{0}$ coincide with each other, while the $Q^{A}_{0}$ and $Q^{P}$ vanish: Note that, on the other hand, $Q^{S}$ and $Q^{P}$ vanish at the ultrarelativistic limit. 

Under the presence of the vectorial density matrices like the $Q^{V}_{i}$, $Q^{A}_{i}$ and $Q^{T}_{\mu\nu}$, the $O(3)$ rotational symmetry will be broken.
For $P\times T$-, $P$-, $T$- and $1$ (no symmetry)-invariant solutions,
simple classification is impossible, because there are various possibilities
of the angular momentum dependences of the $Q$. This situation demands us futher investigation in more detail in the future.
It must be noted that we have to solve QED-HF equations to find which type of the density matrices ( and solutions ) will be realized. The natural orbital analysis of a resulted solution is useful to elucidate the type and magnitude of broken symmetry.

\section{Post QED-HF methods}

\subsection{QED-CASSCF and CASPT2 methods}

The standard approach for inclusion of fluctuation from the HF mean field
approximation is the MBPT approach~[6]. For the purpose, the total Hamiltonian
in (9) is devided into the zeroth-order and interaction parts as
\begin{eqnarray}
\hat{H}_{0} &=& \hat{H}_{K} + \hat{H}^{(HF)}_{I}, \\
\hat{H}_{int} &=& \hat{H}_{I}-\hat{H}^{(HF)}_{I},
\end{eqnarray}
where $\hat{H}^{(HF)}_{I}$ denotes the HF mean field interaction potential.
The remaining interaction $\hat{H}_{int}$ is regarded as the perturbation
to the HF solution in the case of the MBPT approach~[6].

However, such single-reference (SR) MBPT approach often breaks down in the case
of near degenerate ground states, for which generalized QED-HF solutions exist,
because of the HF instability in (27)~[11]. To consider these problems, we introduce the density matrix which is defined as
\begin{equation}
\left(
\begin{array}{cc}
\rho^{++}_{ij}(\kappa) & \rho^{+-}_{ij}(\kappa) \\
\rho^{-+}_{ij}(\kappa) & \rho^{--}_{ij}(\kappa)
\end{array}
\right)= \left(
\begin{array}{cc}
\langle\Phi(\kappa)|\hat{a}^{\dagger}_{j}\hat{a}_{i}|\Phi(\kappa)\rangle &
\langle\Phi(\kappa)|\hat{b}_{j}\hat{a}_{i}|\Phi(\kappa)\rangle  \\
\langle\Phi(\kappa)|\hat{a}^{\dagger}_{j}\hat{b}^{\dagger}_{i}|\Phi(\kappa)\rangle &
\langle\Phi(\kappa)|\hat{b}_{j}\hat{b}^{\dagger}_{i}|\Phi(\kappa)\rangle 
\end{array}
\right),
\end{equation}
where we write $(\alpha, \beta, \beta^{*}, \gamma)$ as $(\kappa)$. 
Then we diagonalize the density matrix of the most stable QED-HF solution for systems under consideration~[14], we obtain the natural orbital given by four-component bispinor
$\eta^{(\pm)}_{i}$ and $m_{i}$ as its occupation number. By using the occupation number $m_{i}$, we can select the active space for treatments of the near degeneracy effects~[18]. Then the trial CI wavefunction for the QED-MCSCF are easily constructed. The QED-MCSCF wavefunction is obtained like the nonrelativistic case: 
\begin{eqnarray}
|\Psi_{0}\rangle &=& \sum_{I}C^{(0)}_{I}e^{i\hat{S}_{1}}|\Phi_{I}\rangle. 
\end{eqnarray}
Here $|\Phi_{I}\rangle$ is the configuration state function ( CSF ),
$C_{I}$ is the CI coefficient, and $e^{i\hat{S}_{1}}$ is the orbital rotation
operator given in previous section. 
With introducing the orthogonal complement to (48), 
$|\Psi_{K}\rangle = \sum_{J}C^{(K)}_{J}|\Phi_{J}\rangle$,
the variational operator of the MC function is given as
\begin{eqnarray}
\hat{T} &=& \sum_{K\neq 0}T_{K}\{|\Psi_{K}\rangle\langle \Psi_{0}|+|\Psi_{0}\rangle\langle \Psi_{K}|\}, 
\end{eqnarray}
where, $\hat{T}^{\dagger}=\hat{T}$ and $\hat{S}^{\dagger}_{1}=\hat{S}_{1}$ will be satisfied. Then we can write the transformed energy as
\begin{eqnarray}
E &=& \langle \Psi_{0}|e^{-i\hat{T}}e^{-i\hat{S}_{1}}\hat{H}e^{i\hat{S}_{1}}e^{i\hat{T}}|\Psi_{0}\rangle.
\end{eqnarray}
This expression is, at a glance, same as the nonrelativistic case. Only the definition of the orbital rotation operator is different. 
From (50), we can formally derive the variational condition such as the generalized Brillouin theorem in our QED theory. 
Therefore the multi-reference ( MR ) MBPT approach based on QED-MCSCF may be feasible like CASPT2~[19] for the nonrelativistic case. 

\subsection{QED-CC and MRCC methods}

The coupled-cluster wave function~[20$\sim$22] can also be introduced to obtain dynamical correlation corrections, by obeying the exponential ansatz:
\begin{eqnarray}
|\Psi_{CC}\rangle &=& e^{i\hat{S}}|\Phi_{0}\rangle. 
\end{eqnarray}
Here the operator $\hat{S}$ is given as
\begin{eqnarray}
\hat{S} &=& \sum_{n}\hat{S}_{n} = \hat{S}_{1}+\hat{S}_{2}+\cdots+\hat{S}_{N}+\cdots. 
\end{eqnarray}
$n$ denotes the excitation level. $|\Phi_{0}\rangle$ is the QED-HF state. 
In the relativistic theory, the expansion given above can not be truncated at N-excitation level, because of the presence of the Dirac sea. Here, the excitation operator $\hat{S}$ is approximately truncated in the second order; $\hat{S} \cong \hat{S}_{1}+\hat{S}_{2}$. Even under this approximation,  $\hat{S}_{2}$ includes the operators given as follows:
\begin{eqnarray}
& & \hat{a}^{\dagger}_{i}\hat{a}_{j}\hat{a}^{\dagger}_{k}\hat{a}_{l}, \qquad \hat{a}^{\dagger}_{i}\hat{a}_{j}\hat{a}^{\dagger}_{k}\hat{b}^{\dagger}_{l}, \qquad \hat{a}^{\dagger}_{i}\hat{a}_{j}\hat{b}_{k}\hat{a}_{l}, \qquad \hat{a}^{\dagger}_{i}\hat{a}_{j}\hat{b}_{k}\hat{b}^{\dagger}_{l}, \qquad  \hat{a}^{\dagger}_{i}\hat{b}^{\dagger}_{j}\hat{a}^{\dagger}_{k}\hat{b}^{\dagger}_{l}, \nonumber \\
& &  \hat{a}^{\dagger}_{i}\hat{b}^{\dagger}_{j}\hat{b}_{k}\hat{a}_{l}, \qquad \hat{a}^{\dagger}_{i}\hat{b}^{\dagger}_{j}\hat{b}_{k}\hat{b}^{\dagger}_{l}, \qquad \hat{b}_{i}\hat{a}_{j}\hat{b}_{k}\hat{a}_{l}, \qquad \hat{b}_{i}\hat{a}_{j}\hat{b}_{k}\hat{b}^{\dagger}_{l}, \qquad \hat{b}_{i}\hat{b}^{\dagger}_{j}\hat{b}_{k}\hat{b}^{\dagger}_{l}.
\end{eqnarray}
With considering that the operators $\hat{a}^{\dagger}\hat{a}$, $\hat{a}^{\dagger}\hat{b}^{\dagger}$, $\hat{b}\hat{a}$, $\hat{b}\hat{b}^{\dagger}$ in $\exp(i\hat{S}_{1})$ form the Lie algebra, there is no other posibility. To write the exponential form, the excitation operators have to be bosonic operators. Thus we have to devide the one-particle function space to two; occupied and unoccupied spaces. Quantum number $i$, $j$, $k$, $l$ will satisfy some conditions like the nonrelativistic case. These operators have to take the form which will give two particle excitation with respect to HF state. The operators given above include redundant operators, which will not give the excitation with operating the HF state. The Schr\"{o}dinger equation is given as
\begin{eqnarray}
\hat{H}e^{i\hat{S}}|\Phi_{0}\rangle &=& E e^{i\hat{S}}|\Phi_{0}\rangle. 
\end{eqnarray}
Then the projected coupled-cluster equations are given as 
\begin{eqnarray}
\langle \Phi_{0}|\hat{H}e^{i\hat{S}}|\Phi_{0}\rangle &=& E_{0}, \\
\langle \Phi_{0}|\hat{S}_{n}\hat{H}e^{i\hat{S}}|\Phi_{0}\rangle &=& E_{0}\langle \Phi_{0}|\hat{S}_{n}e^{i\hat{S}}|\Phi_{0}\rangle.
\end{eqnarray}
If we replace $|\Phi_{0}\rangle$ in (51) with $|\Psi_{0}\rangle$ in (48), we may have the MRCC formulation~[14,23,24]: 
\begin{eqnarray}
|\Psi_{MRCC}\rangle &=& e^{i\hat{S}}|\Psi_{0}\rangle \cong e^{i(\hat{S}_{1}+\hat{S}_{2})}|\Psi_{0}\rangle.
\end{eqnarray}
However, detailed formulations are rather complex because of redandancy. They will be discussed elsewhere. A Schema of this work given in Fig. 2.

\section{Discussion}

In this paper, we have developed our method which combines QED with many-body techniques. We have given the QED Hamiltonian written by creation-annhilation opetrators, and introduced the relativistic Slater determinant in the Thouless form. We have performed the group-theoretical classification of the density matrix. After these preparations, we have discussed the possibilities of the MCSCF and coupled-cluster method in QED. 

Now we discuss the relation between our theory and other relativistic theories.

In the context of the electronic structure of atoms, there are three effects in atoms: The electron correlation, the relativistic effect and the QED effect~[25]. The electronic structure of atoms is determined by the relation of these three factors. The electron correlation depends on the electron numbers and can be treated by several many-body techniques. The relativistic effect becomes large with increasing the atomic number. The relativistic effect in neutral or almost neutral heavy elements can be treated satisfactorily by the Dirac-Coulomb-Breit no-sea scheme~[6], as discussed in introduction. 

In fact, our QED Hamiltonian with neglecting positron states and adopting only the density-density interaction taken into account, will derive the Dirac-Coulomb Hamiltonian~[26,27] ( as illustrated in Fig. 3 ). Futher, we can have the Dirac-Coulomb-Breit Hamiltonian with adding the Breit operator. Moreover, if the small component $\chi$ is approximated by using the large component $\phi$, and take into account the fact that $\chi$ gives small contribution, then the Dirac equation can be reduced to the Schr\"{o}dinger-Pauli equation~[28], for which 2-component spinor $\phi$ ( GSO ) will correctly allow for relativistic effects up to $(v/c)^{2}$. These terms will be calculated by ab initio GSO program package~[4,5].

On the other hand, the case to describe the inner core electrons of heavy elements or, electrons of highly ionized heavy elements such as lithium-like uranium, the QED effect can not be neglected and we must take the Dirac sea into account. This effect can be treated by the perturbation theory in QED~[9]. But this method can only take into account the one-particle QED correction. Practically this method treat the system of atoms with a few electrons. 
For an illminative example of the QED effects, Fig. 4 shows the collision of two uranium atoms~[29]. This collision generates the pair-creations, and the QED effects might be appeared clearly. Moreover, the near orbital degeneracy effects will become important in this case, and also in clusters of these atoms or ions.

In the case of heavy atoms, as the ionicity
becomes high and the electron number decrease, the many-body effect becomes small and the QED effect becomes large. Therefore we propose that our theory should be applied to the cases where both the many-body effects and the QED effects can not be neglected. Heavy elements in middle level of ionicity should be one of the subject for our theory. The Dirac-Coulom-Breit no-sea scheme has been applied to only neutral or almost neutral atoms. The method of the QED correction works only for the highly ionized heavy atoms. Our theory should be suitable for intermediate case between them. The collision of two uranium atoms is also
one of the interesting subject of our theory. Recently, experiments of x-ray
irradiation to cluster plasma are performed, and verious new phenomena were studied~[30]. The cluster of heavy ions under middle revel of ionicity is now 
an interesting subject in this area. The importance of the relativistic and QED effects is discussed in such objects. 
Fig. 5 illustrate scope and limitation of these theories. 
It is noteworthy that many interesting phenomena appear in the intermediate
regime even in the nonrelativistic case~[31]. An application of the QED scheme
to treat the strong electron-electron interaction in solid state physics
is also discussed in relation to spin density wave and charge density wave
states of mixed valence ( MV ) systems~[32].

\acknowledgments

The authors wish to thank Dr. D. Yamaki and Dr. S. Yamanaka for their helpfull discussions, and Mr. Y. Kitagawa for his help to depict the figures. The authors are grateful for the financial support of a Graint-in-aid for Scientific Reserch on Priority Areas ( Nos. 10149105 and 11224209 ) from Ministry of Education, Science, Sports and Culture, Japan.

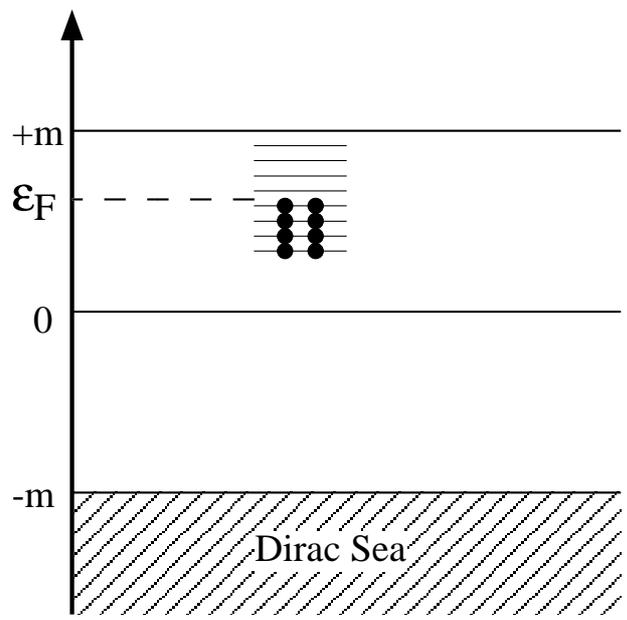
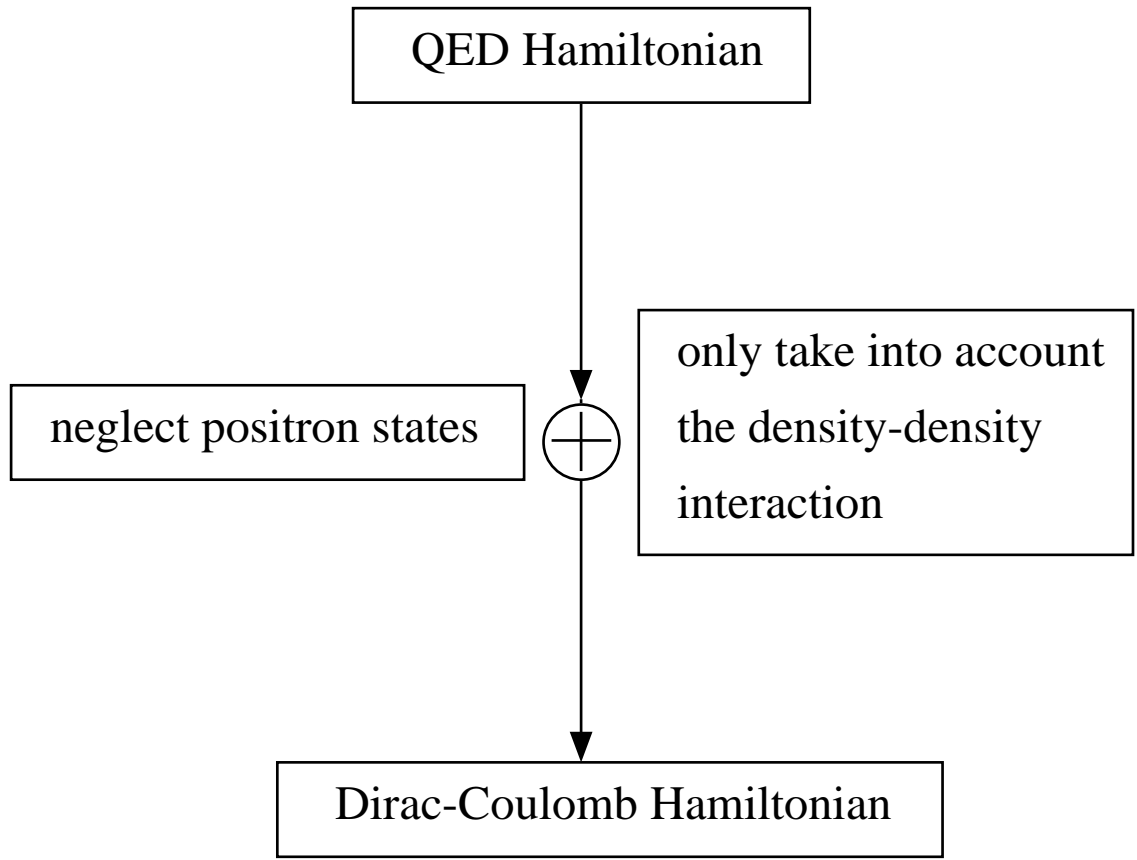
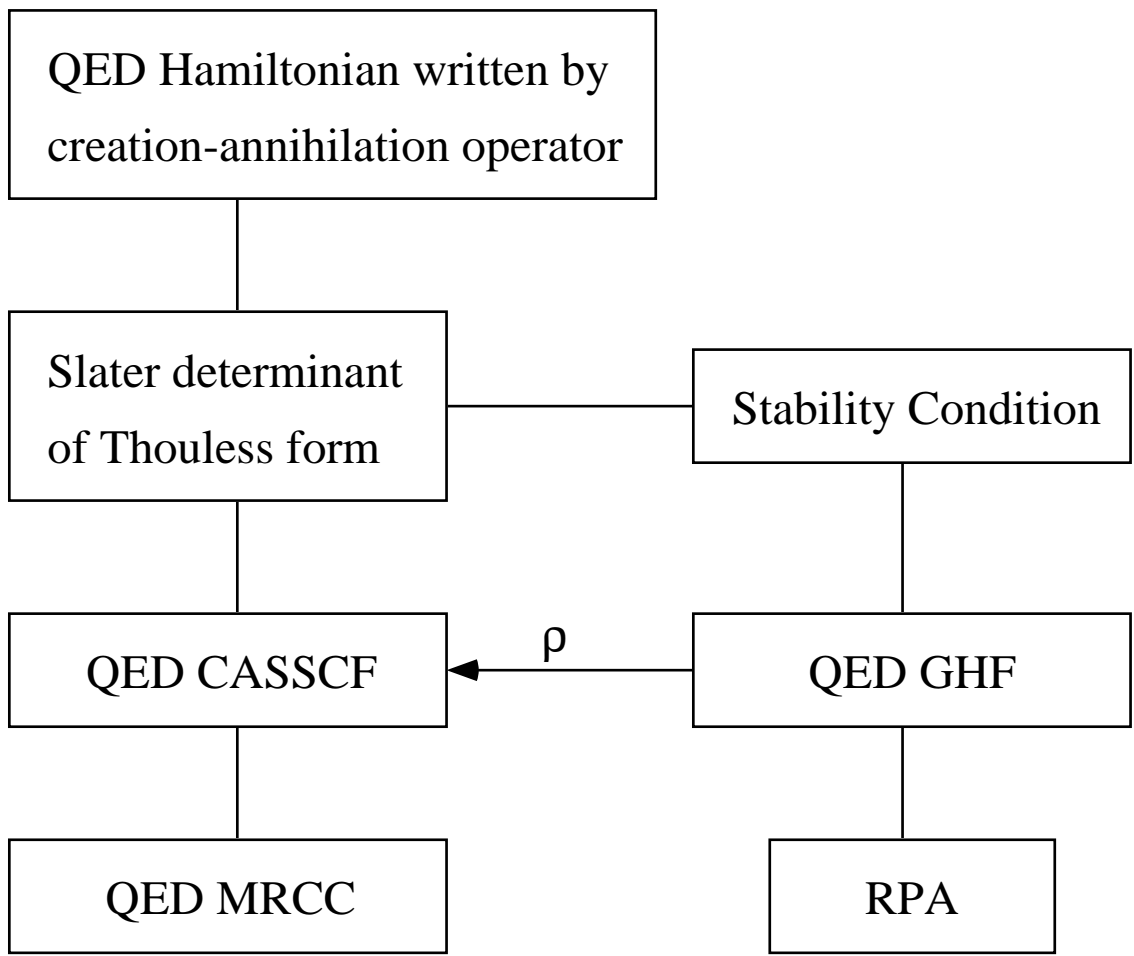
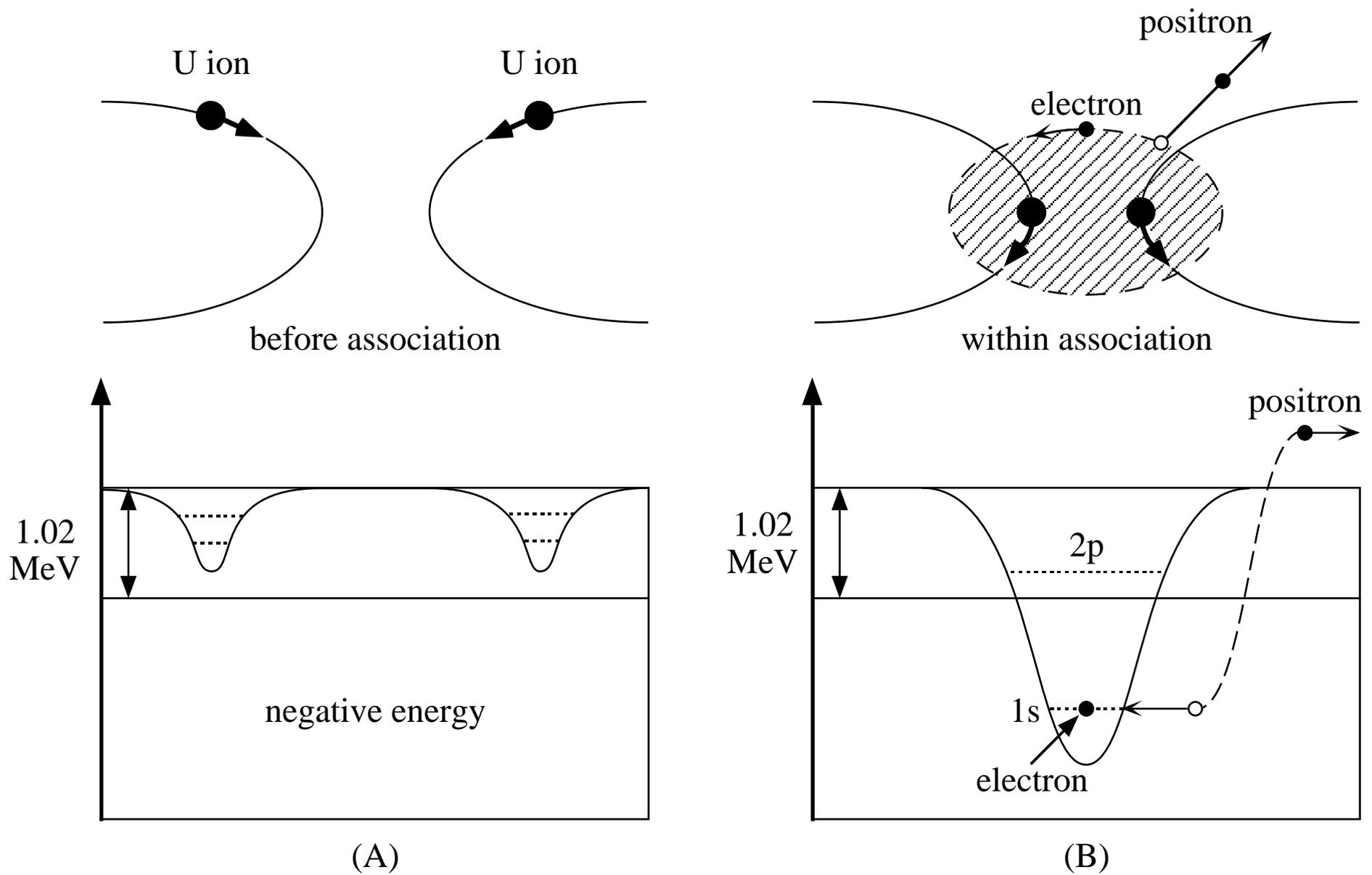
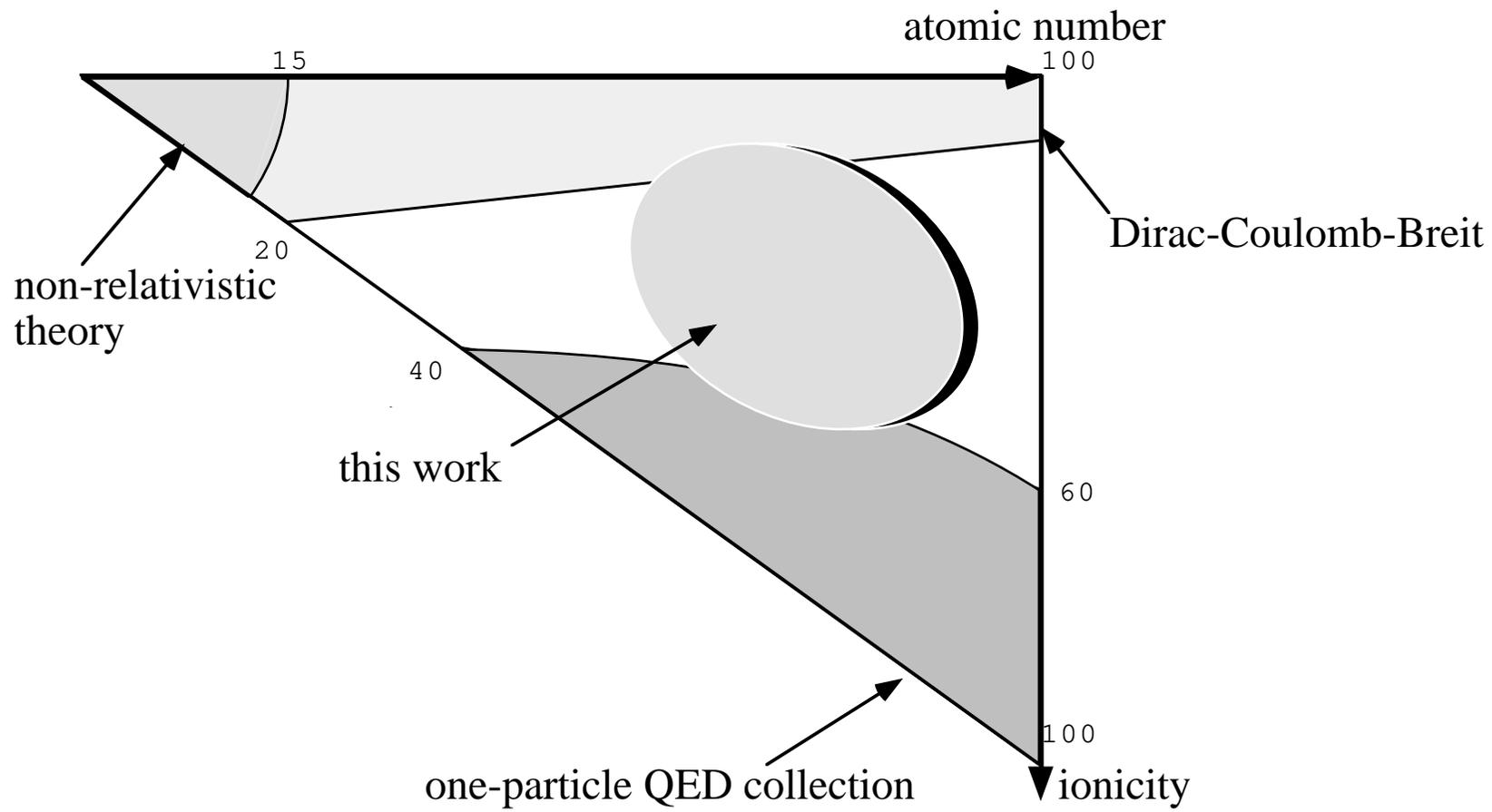
\end{document}